# Streaming supercomputing needs workflow-enabled programming-in-the-large

Justin M Wozniak (presenter)[*], Jonathan Ozik[*], Daniel S Katz[#], Michael Wilde[*]
[*]Argonne National Laboratory, [#]University of Illinois Urbana-Champaign

**Position Statement**: **Simple data analysis applications are common today, but future online supercomputing workloads will need to couple multiple advanced technologies (streams, caches, analysis, and simulations) to rapidly deliver scientific results.** Each of these technologies are active research areas when integrated with high-performance computing. These components will interact in complex ways, therefore coupling them needs to be programmed. **Programming in the large, on top of existing applications, enables us to build much more capable applications and to productively manage this complexity.**

Data analysis applications are growing more complicated, but are still often implemented one step at a time. On a single computer, these steps are linked together by glue code (today, often in Python). However, current and near future hardware platforms offer more parallelism than ever before, and these single applications have difficulty scaling to efficiently use all of it. Another layer of parallelism that is more coarse-grained is an efficient method to use this parallelism, and also is efficient in terms of programmer productivity, allowing the programmer to divide and conquer the programming task, not just the data analysis. We propose highly programmable, implicitly parallel workflow systems to address this challenge.

Science rarely wants to solve a single problem once. Often, scientists need to explore a space, either in parallel or in sequence or both. For example, different machine learning algorithms might be used to find out which is best, or different parameters within one algorithm, or different types of training. In all of these cases, workflows can be used to run the various elements concurrently. Ideally, these parameter studies would be automated with advanced algorithms that drive the workflow.[1]

We believe that Swift (http://swift-lang.org) is a useful solution to this challenge, as it is a full parallel scripting language that allows arbitrarily complex programs, composed of smaller elements, to be written and efficiently executed. These elements (data analysis steps) can be executables or functions, written in a variety of languages, from C and Fortran to C++ and Python and R. The executables can be complex simulations, such as parallel MPI models. And the outer program can be simple, such as a fork-join model that starts a bunch of parallel instances of a data analysis routine on parts of a dataset then combines the results, or complex, such as an optimization or active learning algorithm.

**A remaining challenge: Processing streaming data**. While a handful of task-oriented programming systems (including Swift/T, Charm++, Legion, Chapel) have emerged for high-performance computing, none were originally designed to process streaming data. Data streams from outside the HPC complex are still difficult to connect to compute jobs. High-quality data stream technologies that can initiate and maintain streaming connections of varying performance behavior (e.g., high bandwidth, highly fragmented, bursty) must be developed. In some cases, these need to populate compute-node resident caches for repeated access. Given these techniques (that may be borrowed from the JVM or Python communities), the streams must be connected to tasks emerging from complex workflows as part of, for example, model fitting or data assimilation algorithms. Advanced hierarchical programming models must be provided for computational scientists to integrate and develop high-quality solutions to the scientific problems that drive these workflows.

---

[1] http://www.mcs.anl.gov/~emews/tutorial